\documentclass[11pt,acmsmall,nonacm]{acmart}

\renewcommand\footnotetextcopyrightpermission[1]{}

\usepackage[utf8]{inputenc}
\usepackage[T1]{fontenc}
\usepackage{amsmath}
\usepackage{graphicx}
\usepackage{booktabs}
\usepackage{hyperref}
\usepackage[inline]{enumitem}
\usepackage{subcaption}
\usepackage{adjustbox}
\title{Exploring Topologies in Quantum Annealing: A Hardware-Aware Perspective}

\author{Mario Bifulco}
\affiliation{%
  \institution{University of Turin}
  \city{Turin}
  \country{Italy}}
\email{mario.bifulco@unito.it}

\author{Luca Roversi}
\affiliation{%
  \institution{University of Turin}
  \city{Turin}
  \country{Italy}}
\email{luca.roversi@unito.it}

\begin{document}

\begin{abstract}
  Quantum Annealing (QA) offers a promising framework for solving NP-hard optimization problems, but its effectiveness is constrained by the topology of the underlying quantum hardware. Solving an optimization problem $P$ via QA involves a hardware-aware circuit compilation which requires representing $P$ as a graph $G_P$ and embedding it into the hardware connectivity graph $G_Q$ that defines how qubits connect to each other in a QA-based quantum processing unit (QPU).

  Minor Embedding (ME) is a possible operational form of this hardware-aware compilation. ME heuristically builds a map that associates each node of $G_P$ --- the logical variables of $P$ --- to a chain of adjacent nodes in $G_Q$ by means of one of its minors, so that the arcs of $G_P$ are preserved as physical connections among qubits in $G_Q$.

  The static topology of hardwired qubits can clearly lead to inefficient compilations because $G_Q$ cannot be a clique, currently. We propose a methodology and a set of criteria to evaluate how the hardware topology $G_Q$ can negatively affect the embedded problem, thus making the quantum optimization more sensible to noise.

  We evaluate the result of ME across two QPU topologies: Zephyr graphs (used in current D-Wave systems) and Havel-Hakimi graphs, which allow controlled variation of the average node degree. This enables us to study how the ratio `number of nodes/number of incident arcs per node' affects ME success rates to map $G_P$ into a minor of $G_Q$.

  Our findings, obtained through ME executed on classical, i.e. non-quantum, architectures, suggest that Havel-Hakimi-based topologies, on average, require shorter qubit chains in the minor of $G_P$, exhibiting smoother scaling of the largest embeddable $G_P$ as the QPU size increases. These characteristics indicate their potential as alternative designs for QA-based QPUs.
\end{abstract}

\maketitle

\section{Introduction}

\subsection{Quantum annealing in the landscape of quantum computing}

In the current landscape of quantum computing, two principal paradigms can be identified.
The first is \emph{digital quantum computing}, which relies on quantum gates—the quantum counterpart of classical logic gates—and offers universality at the cost of high control and error-correction overhead.
The second is \emph{analog quantum computing}, typically realized through \emph{Adiabatic Quantum Computing} (AQC)~\cite{AQC}.
Although AQC is theoretically universal, no large-scale universal implementation has yet been demonstrated.
A practically relevant subset of AQC is \emph{Quantum Annealing} (QA)~\cite{QA}, a non-universal but effective approach for solving NP-hard optimization problems by exploiting quantum tunneling to explore complex energy landscapes.

\subsection{Workflow and compilation challenges}

The QA workflow can be schematically divided into three main stages:
\begin{enumerate*}
  \item Express the optimization problem $P$ in \emph{Quadratic Unconstrained Binary Optimization} (QUBO) form~\cite{lucas}, which entails representing $P$ as a weighted graph $G_P$, often a clique;
  \item Embed $G_P$ into the hardware graph $G_Q$ that encodes the physical topology of the quantum processing unit (QPU)~\cite{ME, MEdwave, MENP, localembedding};
  \item Execute the annealing process, during which the physical system described by $G_Q$ evolves toward a low-energy state encoding the solution of $P$.
\end{enumerate*}

The second stage is the crucial hardware-aware compilation step, known as \emph{Minor Embedding} (ME), which effectively routes the logical problem's connectivity across the physical QPU topology.
Since the structure of $G_Q$ constrains which problems can be efficiently embedded, the quality of this stage plays a key role in determining the performance of the overall computation.

\paragraph{Problem formulation.}
Given a logical problem graph $G_P$ and a hardware topology $G_Q$, the compilation goal can be expressed as:
\emph{minimize the total qubit chain length required to embed $G_P$ into $G_Q$, while preserving all logical adjacency.}
This formulation clarifies the connection between embedding efficiency and hardware-aware optimization.

\paragraph{Minor embedding.}
A graph $G_m$ is a \emph{minor} of another graph $G$ if it can be obtained through edge contractions and vertex or edge deletions.
In QA, the objective is to find a minor $G_m$ of $G_Q$ that is isomorphic to $G_P$, effectively mapping logical variables to physical qubits through chains.
Long qubit chains require stronger coupling penalties to maintain coherence, making the system more fragile and noise-sensitive.

\subsection{Topological limitations and motivation for this study}

Recent generations of D-Wave annealers illustrate this trade-off between connectivity and scalability.
The Pegasus topology~\cite{Pegasus}, for instance, allows embeddings of up to roughly 100-variable cliques on a QPU with more than 5000 qubits~\cite{mscthesis}.
The more recent Zephyr topology, while featuring fewer physical qubits (around 4000), provides denser connectivity and can embed cliques up to approximately 150-variable cliques.
Such differences highlight the crucial role of topology in determining the expressive power of a quantum annealer, to us, ``expressive power'', in this context, meaning ``the maximal dimension of a clique that can be minor embedded into an annealer.''
This observation motivates a systematic study of how different QA-based QPUs topologies influence the embedded problem $G_E$.

\section{QPU Topology}

The topology can be analyzed at both coarse and fine granularity.
At a coarse level, metrics such as the number of nodes, edges, and the resulting average degree provide a first approximation of how densely connected the QPU is.
At a finer level, metrics such as \emph{regularity} and \emph{modularity} capture local and global patterns affecting embedding quality.
In particular, \emph{regularity} quantifies how uniform the degree distribution is across nodes and \emph{modularity} measures the extent to which the graph can be partitioned into clusters with dense internal connections and sparse external ones.

\begin{figure}
  \centering
  \begin{subfigure}{0.48\textwidth}
    \centering
    \includegraphics[width=\linewidth]{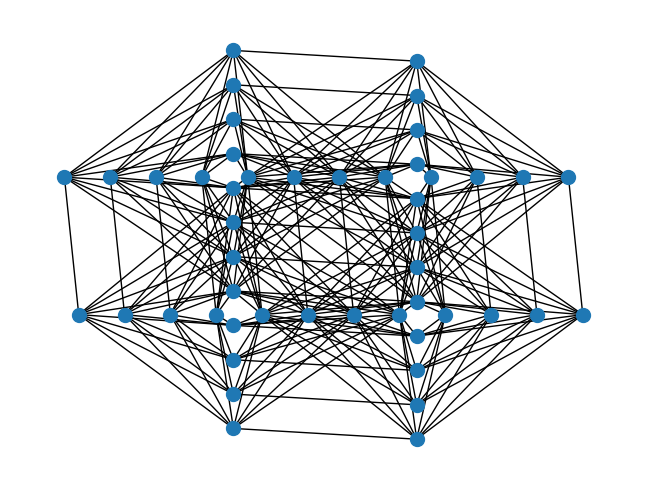}
    \caption{Zephyr QPU topology $G_Q$ with parameters $m=1$, $t=4$.}
    \label{fig:zephyr_qpu}
  \end{subfigure}\hfill
  \begin{subfigure}{0.48\textwidth}
    \centering
    \includegraphics[width=\linewidth]{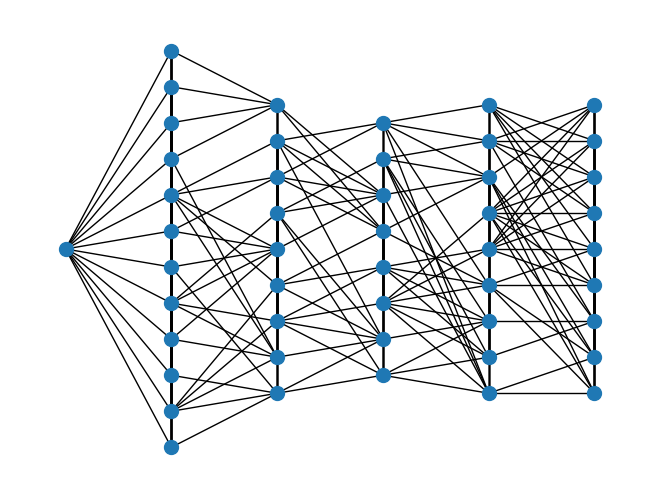}
    \caption{Havel-Hakimi QPU topology $G_Q$: 48 nodes of degree 12.}
    \label{fig:hh_qpu}
  \end{subfigure}
  \caption{QPUs with comparable size and average degree but distinct structural properties.}
  \label{fig:zephyr_hh}
\end{figure}

We use these fine-grained metrics to compare Zephyr and Havel-Hakimi graph topologies.
Figure~\ref{fig:zephyr_hh} provides a pictorial comparison of the two, for graphs of comparable size and average degree.

\subsection{Zephyr Graph}

The Zephyr graph~\cite{QPU} defines the current D-Wave QPU topology.
It is parameterized by grid size $m$ and tile size $t$, forming an $m \times m$ array of cross-shaped tiles, each composed of $t$ qubits in each cross arm.
Each tile represents a local connectivity unit (the ``cross''), while the grid defines how these tiles are interconnected to form the global QPU layout.
Figure~\ref{fig:zephyr_qpu} shows an example of a Zephyr topology for parameters $m=1$ and $t=4$.

We explored configurations with $(m,t) \in [2,7] \times [1,25]$, generating 150 QPU graphs ranging from 40 to 10,500 nodes and from 114 to 508,750 edges.

\subsection{Havel-Hakimi Graph}

Havel-Hakimi graphs are constructed by specifying the uniform node degree $deg$ and the total number of nodes $num\_qubits$.
The Havel-Hakimi algorithm~\cite{hh} iteratively connects nodes following a descending degree sequence.
Havel-Hakimi graphs are more regular than Zephyr ones but, they lack modularity, since there is no a priori structure repeated periodically.
We generated regular instances with $deg \in \{5+25k \mid k=0,\dots,4\}$ and $num\_qubits \in \{50+350m \mid m=0,\dots,29\}$, yielding 150 graphs ranging from 50 to 10,200 nodes and from 125 to 535,500 edges.
These parameters were chosen so that Havel-Hakimi graphs cover approximately the same ranges of size and average degree as the Zephyr graphs, allowing a fair comparison in terms of expressive power.

To ensure physical plausibility, we imposed a maximum node degree of about 100.
This value reflects a reasonable upper bound based on foreseeable fabrication constraints in near-term QPU architectures: for reference, current Zephyr topologies reach node degrees up to 20 and in 10 years D-Wave only succeed to double the degree from the original Chimera topology.
This constraint allows us to explore the scalability of connectivity while remaining within a range that could be physically realizable in future QPUs.

\section{Methodology}

We generated Zephyr and Havel-Hakimi graphs, standing for $G_Q$s, using \texttt{zephyr\_graph} from \texttt{dwave\_networkx}\footnote{\url{https://docs.dwavequantum.com/en/latest/ocean/api_ref_dnx/}} and \texttt{havel\_hakimi\_graph} from \texttt{networkx}\footnote{\url{https://networkx.org/}}.
For each $G_Q$, we determined the largest embeddable clique $G_P = K_n$ using the \texttt{find\_embedding} function from \texttt{minorminer}\footnote{\url{https://docs.dwavequantum.com/en/latest/ocean/api_ref_minorminer/source/index.html}}.
The resulting embedded graph $G_E$ associates each node of $G_P$ with a chain of adjacent qubits in $G_Q$, preserving the logical adjacency structure.

We use $K_n$ cliques as benchmark because real-world QUBO instances $G_P$ often are highly connected graphs.
Fixed $G_Q$, for each embedding of $K_n$ generating $G_E^{K_n}$, we collect:
\begin{itemize}
  \item the \emph{QPU descriptors}: \emph{type}, \emph{parameters}, \emph{number of nodes} and \emph{edges}, \emph{average degree} $\tilde{d}_n$ of $G_Q$;
  \item the \emph{descriptive statistics}: \emph{mean}, \emph{median}, and \emph{mode} of qubit chain lengths in $G_E^{K_n}$.
\end{itemize}
Moreover, relatively to every $G_Q$, we record the size $\operatorname{\textit{max}}$ of the largest $K_{\operatorname{\textit{max}}}$ that ME can embed into $G_Q$, generating $G_E^{K_{\operatorname{\textit{max}}}}$.

Finally, for every $G_Q$, we normalize every average degree $\tilde{d}_n$, and the value $\operatorname{\textit{max}}$ by the number of qubits in $G_Q$.
The normalization provides a unified comparison space to highlight structural correlations between QPU topologies and embedding performance, because it removes the scale dependence on QPU size, allowing us to compare topologies with different numbers of qubits on equal footing.
The normalized quantities described above provide the basis for evaluating how different QPU topologies affect both embedding capacity and chain compactness, as reported in the following section.

\section{Results}

\begin{figure}
  \centering
  \begin{subfigure}{0.48\textwidth}
    \centering
    \includegraphics[width=\linewidth]{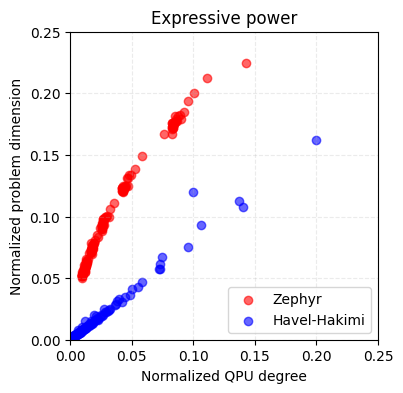}
    \caption{Normalized size of embedded $K_{\operatorname{\textit{max}}}$ across Zephyr and Havel-Hakimi QPU $G_Q$.}
    \label{fig:zephyr_vs_hh}
  \end{subfigure}\hfill
  \begin{subfigure}{0.47\textwidth}
    \centering
    \includegraphics[width=\linewidth]{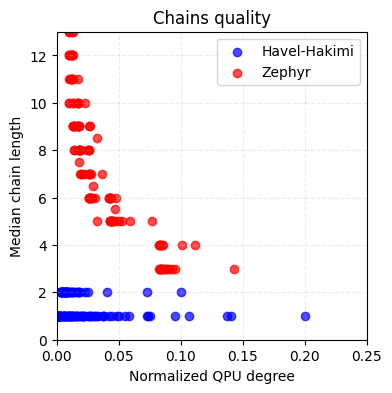}
    \caption{Median chain length of $G_E^{K_{\operatorname{\textit{max}}}}$ for every Zephyr and Havel-Hakimi QPU $G_Q$}
    \label{fig:chains}
  \end{subfigure}
  \caption{The largest dimension (a) and median chain length (b) of $G_E^{K_{\operatorname{\textit{max}}}}$ for Zephyr and Havel-Hakimi QPU topologies.}
  \label{fig:zephyr_vs_hh and chains}
\end{figure}

Figure~\ref{fig:zephyr_vs_hh} summarizes the results regarding the maximum size clique embeddable in a given QPU.
The horizontal axis represents the normalized average degree of $G_Q$, while the vertical axis reports the normalized size of the largest clique $G_P$ successfully embedded, i.e. $G_E^{K_{\operatorname{\textit{max}}}}$.

Havel-Hakimi graphs (blue points) display an almost linear scaling: as connectivity increases, the size of the embeddable clique grows proportionally.
Zephyr topologies (red points) exhibit a sublinear trend—highly expressive at low degrees but saturating as connectivity increases.

Figure~\ref{fig:chains} shows that embeddings on Havel-Hakimi graphs yield shorter qubit chains and smoother scaling.
Overall, these results suggest that synthetic topologies with controlled degree distributions can enhance embedding capacity without excessive chain length, thus improving robustness and scalability of solving problems by QA.

\section{Future work}

Future work will extend this comparative analysis to a broader set of synthetic and physically inspired QPU topologies $G_Q$.
We also plan to investigate correlations between structural indicators of $G_Q$ and practical embedding quality metrics, such as chain stability and noise susceptibility in $G_E$.
As a further direction, we intend to explore new candidate topologies derived from graph-theoretical principles, focusing on structures that could minimize the length of qubit chains while maintaining feasible connectivity and manufacturability.
These studies aim to support the co-design of hardware architectures and embedding heuristics, promoting QPU topologies that balance physical realizability, connectivity, and embedding efficiency.

\bibliographystyle{ACM-Reference-Format}
\bibliography{ref}

\end{document}